\documentstyle[prl,twocolumn,aps,epsf]{revtex}

\begin{document}
\title{Cryptoferromagnetic state in superconductor-ferromagnet multilayers}
\author{F.S. Bergeret$^{(1)}$, K.B. Efetov$^{(1,2)}$, A.I. Larkin$%
^{(3,2,1)} $}
\address{$^{(1)}$Theoretische Physik III,\\
Ruhr-Universit\"{a}t Bochum, 44780 Bochum, Germany\\
$^{(2)}$L.D. Landau Institute for Theoretical Physics, 117940 Moscow, Russia 
\\
$^{(3)}$Theoretical Physics Institute, University of\\
Minnesota, Minneapolis, MN 55455, USA}
\maketitle

\begin{abstract}
We study a possibility of a non-homogeneous magnetic order
(cryptoferromagnetic state) in heterostructures consisting of a bulk
superconductor and a ferromagnetic thin layer that can be due to the influence
of the superconductor. The exchange field in the ferromagnet may be strong
and exceed the inverse mean free time. A new approach based on solving the
Eilenberger equations in the ferromagnet and the Usadel equations in the
superconductor is developed. We derive a phase diagram between the
cryptoferromagnetic and ferromagnetic states and discuss the
possibility of an experimental observation of the CF state in different
materials.

\end{abstract}


In the last years, the interest in experiments on
superconducting-ferromagnet ($S/F$) hybrid structures has grown rapidly.
Such structures show the coexistence of these two antagonistic orderings but
their mutual influence is still a controversial point \cite
{koorevaar,strunk,jiang,zabel,verbanck,aarts}. In these experiments, the
multilayers contained strong ferromagnets like $Fe$ or $Gd$ with the Curie
temperature up to $1000K$ and superconductors with transition
temperatures not exceeding $10K$, like $Nb$ or $V$. 

Naturally, in most theoretical works only the influence of the ferromagnet
on the superconductivity of $S/F$ systems was considered \cite
{radovic,sipr,melo}. One may argue that a modification of the magnetic
ordering would need energies of the order of the Curie, which is much larger than the superconducting transition
temperature $T_{c}$. Therefore, any change of the ferromagnetic order would
be less energetically favorable than the destruction of the
superconductivity in the vicinity of the ferromagnet.

This simple argument was questioned in a recent experimental work \cite
{muhge}, where $Nb/Fe$ bilayers were studied using different experimental
techniques. Direct measurements using the ferromagnetic resonance 
showed that in several samples with thin ferromagnetic layers $(10- 15$%
\AA $)$ the average magnetic moment started to decay at the superconducting
transition temperature $T_{c}$. The measurements were possible only in a
limited range of the temperatures below $T_{c}$ and the decrease of the
magnetic moment in this interval reached $10\%$ without any sign of a
saturation. As a possible explanation of the effect, it was assumed in Ref. \cite{muhge} that the superconductivity affected the magnetic order causing a
domain-like structure.

A possibility of a domain-like magnetic structure in presence of
superconductivity has been first suggested by Anderson and Suhl long ago 
\cite{suhl}. They argued that a weak ferromagnetism of localized electrons
should not destroy the superconductivity in the conduction band. Instead, it
may become more favorable energetically to build a domain structure called  {\em cryptoferromagnetic state} \cite{suhl}. Later this state was
investigated  both theoretically and experimentally in  detail (for
review see, e.g.\cite{review}).

In this paper, we investigate theoretically the possibility of a
cryptoferromagnetic-like (CF) state in $S/F$ bilayers with parameters
corresponding to the structures used in the experiments \cite
{koorevaar,strunk,jiang,zabel,verbanck,aarts,muhge}. Such a study is very
important because it may allow to clarify the question about the
cryptoferromagnetic state in the experiment \cite{muhge} and to make
predictions for  other $S/F$ multilayers. From the theoretical point of
view, large magnetic energies involved make the problem quite non-trivial
and demand development of new approaches.

To the best of our knowledge the possibility of a non-homogeneous magnetic
order in multilayers was considered only in Ref. \cite{buzdin}. However,
although the authors of Ref. \cite{buzdin} came to the conclusion that the
domain-like structure due to the interaction with the superconductor was
possible, the results obtained can hardly be used for quantitative
estimates. For example, they assumed that the period of the structure $b$
had to be not only  much smaller than  the size of the Cooper pair $\bar{\xi}$%
, but also than  $\bar{\xi}\sqrt{T_{c}/h}$, where $h$ is the energy of interaction
of conduction electrons (CEs) with the localized magnetic moments (LMs). In addition, a very rough
boundary condition at the $S/F$ boundary was used.

In contrast, we present here a microscopic derivation of the phase diagram
valid for realistic parameters of the problem involved. We will show that
the phase transition between the CF and ferromagnetic (F) phases is
continuous and the period of the structure $b$ goes to infinity at the
critical point. The only restrictions we use  are 
\begin{equation}
d\ll \xi _{F}=v_{0}/h,\qquad T_{c}\ll h\ll \epsilon _{0}  \label{a1}
\end{equation}
where $d$ is the thickness of the ferromagnetic layer, $v_{0}$ and $%
\varepsilon _{0}$ are the Fermi-velocity and Fermi-energy.

Even in the such strong ferromagnet as iron, $\xi _{F}$ is of the order $10$%
\AA . For weaker ferromagnets like $Gd$, $\xi _{F\text{ \ }}$is considerably
larger and the inequalities (\ref{a1}) can be fulfilled rather easily.

We assume that the superconductor occupies the half-space $x>0$ while the
ferromagnetic film occupies the region $-d<x<0$ and write the Hamiltonian as 
\begin{equation}
H\!\!\!=\!H_{BCS}+\!\!\gamma \!\!\int d{\bf r}\Psi _{\alpha }^{+}({\bf r})\left[ 
{\bf h}({\bf r}){\bf \sigma }\right] _{\alpha \beta }\Psi _{\beta }({\bf r}%
)+H_{M}  \label{hamiltonian}
\end{equation}
where $H_{BCS}$ is the usual BCS Hamiltonian (in the presence of
non-magnetic impurities) describing the superconducting state in the $S$
layer, $\gamma$ is a constant which will be put to 1 at the end. The second term in Eq. (\ref{hamiltonian}) stands for the
interaction between the LMs of the ferromagnet and the
 CEs, where $\!{\bf h}$ is the exchange field and $%
{\bf \sigma }$ is the vector containing the Pauli matrices as components. We neglect the influence of the
LMs on the orbital motion of the CEs since the exchange interaction is the
dominant Cooper pair breaking mechanism \cite{review} for the problem
involved. The term $H_{M}$ describes the interaction between the LM  in the ferromagnet.

Our aim is to obtain an expression for the free
energy of the system for different magnetic structures in the F layer. To
determine the contribution of an inhomogeneous alignment of magnetic spins
to the total energy we use the limit of a {\em continuous} material and
replace the spins by classical vectors. We assume that the anisotropy energy
of the ferromagnet is smaller than the exchange energy and hence there is no
easy axis of magnetization. This can definitely be a good approximation for
iron with a cubic lattice used in the work \cite{muhge}. The energy $H_{M}$
of a non-homogeneous structure can be written in the continuum limit as 
\begin{equation}
H_{M}=\int J\left[ \left( {\bf \nabla }S_{x}\right) ^{2}+\left( {\bf \nabla }%
S_{y}\right) ^{2}+\left( {\bf \nabla }S_{z}\right) ^{2}\right] dV,
\label{exchange-energy}
\end{equation}
where the magnetic stiffness $J$ characterizes the strength of the coupling
between LMs in the F layer and $S_{i}$'s are the components of a unit vector. Writing ${\bf S}=\left( 0,-\sin \Theta , \cos \Theta \right)$
and minimizing the energy $H_{M}$ we obtain the equation $\Delta \Theta=0$. We consider only the solutions of this equation that are of interest for us:
\begin{equation}
a)\text{ }\Theta =0,\text{ \ \ \ \ \ \ }b)\text{\ }\Theta =Qy   \label{a3}
\end{equation}
The solution a) in Eq. (\ref{a3}) corresponds to the F state, whereas the solution b) describes  a  CF state with a homogeneously rotating magnetic moment. The wave
vector of this rotation is denoted by $Q$. The magnetization is chosen to be
parallel to the FS interface, {\em i.e}. to the $yz-$plane. This allows to
neglect Meissner currents in the superconductor. With all this assumptions the 
 magnetic energy $\Omega _{M}$  (per unit surface area)
is given by 
\begin{equation}
\Omega _{M}=JdQ^{2}  \label{magnetic}
\end{equation}
The corresponding energy of the F state equals zero.

The superconducting part of the energy can be calculated deriving from Eq. (%
\ref{hamiltonian}) proper Eilenberger equations \cite{eilen} for the
superconductor and the ferromagnet, solving these equations and then
matching the solutions. In practice, this is difficult and we simplify the
problem considering the ``dirty limit'' $l\ll \xi _{0}$, where $l$ is the
mean free path and $\xi _{0}=v/T_{c}$ is the coherence length of the
superconductor in the clean limit, which allows to use the more simple Usadel
equations \cite{usadel}. If we assume that $|\tau |\ll 1$, $\tau =\left(
T-T_{c}\right) /T_{c}$, the Usadel equations together with the
self-consistency equation can be further reduced to the Ginzburg-Landau (GL)
equation\cite{agd,degennes,abrikosov}. However, the latter equation can be
used only sufficiently far from the $S/F$ boundary at distances exceeding $%
\bar{\xi}\sim \sqrt{\xi _{0}l}$. At the distances of the order of $\bar{\xi}$
one should write again the Usadel equations but in the limit $|\tau |\ll 1$
they can be linearized. This is a conventional scheme of calculation for
interfaces between superconductors and normal metals or ferromagnets.

Writing the Usadel equations in the ferromagnet may not be  a good
approximation because the exchange energy $h$ in realistic cases is not
necessarily smaller than $1/\tau _{tr}$, where $\tau _{tr}$ the mean free
time, and so one should write in this region the Eilenberger equations. At
the end one should match the solutions of all the equations.

Now we start the  calculations following this program. The loss of the
superconducting energy due to the suppression of the superconductivity in
the $S$-layer can be found from the solution of the GL equation for the
order parameter $\Delta ({\bf r})$. At distances $x\gg \bar{\xi}$, the
proper solution is \cite{agd,degennes,abrikosov} 
\begin{equation}
\Delta (x)=\Delta (T)\tanh \left( \frac{x}{\sqrt{2}\xi (T)}+C\right) ,
\label{orderparameter}
\end{equation}
where $\Delta (T)=\sqrt{\frac{8\pi ^{2}}{7\zeta (3)}|\tau |}T_{c}\equiv
\Delta _{0}\tau ^{1/2}$ is the value of the order parameter in the bulk
superconductor, $\xi (T)=\sqrt{\frac{\pi D}{8T_{c}}}|\tau |^{-1/2}$ is the
characteristic scale of the spacial variation of $\Delta \left( {\bf r}%
\right) $, $D$ is the diffusion coefficient in the superconductor, and $C$
is a constant. Substituting $\Delta \left( x\right) $, Eq. (\ref
{orderparameter}), into the GL free energy functional one can evaluate the
loss of the superconducting energy at the $F/S$ interface per unit surface
area as function of $C$\cite{degennes} 
\begin{equation}
\Omega _{S}=\frac{\sqrt{\pi }}{6\sqrt{2}}|\tau |^{3/2}\left( 2+K\right)
(1-K)^{2}  \label{energysuper}
\end{equation}
where $K=\tanh C$. The influence of the ferromagnet on the superconductivity
is determined by the parameter $K$ that will be found by minimizing the
total energy.

The contribution $\Omega _{M/S}$ of the second term in (\ref{hamiltonian})
to the total energy has still to be determined. First, we write the
Eilenberger equation for the magnetic moment ${\bf h}\left( {\bf r}\right) $
depending on coordinates. Introducing the quasiclassical matrix Green
function $\check{g}_{\omega }({\bf r},{\bf p}_{0})$ 
\[
\check{g}=\left( 
\begin{array}{cc}
\hat{g} & -\hat{f} \\ 
\hat{f}^{+} & -\hat{g}^{+}
\end{array}
\right) 
\]
one derives in the standard way the Eilenberger equation in the spin$\otimes 
$particle-hole space 
\begin{equation}
\left[ \left\{ \omega \check{\tau}_{3}-i\check{\Delta}+i\gamma \check{V}+i%
\check{\Sigma}{\it _{imp}}\right\} ,\check{g}\right] +{\bf v}_{0}\nabla _{%
{\bf r}}\check{g}=0\,.  \label{eilenbergercompleta}
\end{equation}
where ${\bf p}_{0}$ and ${\bf v}_{0}$ are the momentum and velocity at the
Fermi-surface.

In Eq. (\ref{eilenbergercompleta}), $\check{\tau}_{i}$, $i=1,2,3$, are Pauli
matrices in the particle-hole $\!$ space, $\check{\Delta}\!\!=\!\!\check{\tau}_{1}\otimes
i\sigma _{y}\Delta ({\bf r})$, $\check{V}={\rm Re}\left( {\bf h}({\bf r})%
{\bf \sigma }\right) \otimes \check{1}+{\rm Im}\left( {\bf h}({\bf r}){\bf %
\sigma }\right) \otimes \check{\tau}_{3}$, and $\Delta $ should be
determined self-consistently 
\begin{equation}
\Delta ({\bf r})=-\frac{i}{2}\pi \nu \lambda _{0}T\sum_{n}<f_{12}({\bf r},%
{\bf p}_{0},\omega _{n})>_{0}\,,  \label{selbstconsist}
\end{equation}
where $\left\langle ...\right\rangle _{0}$ denotes averaging over the Fermi
velocity and $\lambda _{0}$ is the constant of the electron-electron
interaction, $\nu $ is the density of states. We assume for simplicity that $%
\lambda _{0}=0$ and hence $\Delta =0$ in the ferromagnet. At the same time, $%
h=0$ in the superconductor. The term $i\check{\Sigma}_{{\it imp}}$ describes
scattering by impurities. For a short range interaction, $\check{\Sigma}_{%
{\it imp}}=-\frac{i}{2\tau }\left\langle \check{g}\right\rangle _{0}$. Eq. (%
\ref{eilenbergercompleta}) is complemented by the normalization condition $%
\check{g}^{2}=\check{1}$. Once we know $\hat{g}$, we can determine $\Omega
_{M/S}$ using the expression \cite{agd}: 
\begin{equation}
\Omega _{M/S}=-i\pi T\nu _{0}\sum_{\omega }\int_{0}^{1}d\gamma \int d^{3}%
{\bf r}({\bf h\sigma })_{\alpha \beta }\left\langle g_{\beta \alpha
}\right\rangle _{0}\,  \label{energiaMS}
\end{equation}
Near $T_{c}$, the anomalous functions $\hat{f}$ and $\hat{f}^{+}$ are small
and $\hat{g}\approx sgn\left( \omega \right) $. Then, in the limit $T_{c}\ll
h$ the off-diagonal component (1,2) in particle-hole space of the equation (%
\ref{eilenbergercompleta}) in the region $-d<x<0$ is

\begin{eqnarray}
{\bf v}_{0}\nabla \hat{f}\!\! &=&\!\!-i\hat{V}\hat{f}^{(F)}+i\hat{f}^{(F)}%
\hat{V}^{\ast }\!\!-\frac{sgn\left( \omega \right) }{\tau }(\hat{f}^{(F)}-<\!%
\hat{f}^{(F)}\!>)  \label{eilenberger12} \\
\hat{V} &=&h(x)\sigma _{z}\exp (iQy\sigma _{x})  \nonumber
\end{eqnarray}
$h$ is the strength of the exchange field in the $F$-layer.

Assuming that $d\ll v_{0}/h$ we can relate the values of the function $\hat{f%
}^{(F)}({\bf v}_{{\bf 0}},{\bf r})$ at the interface, {\em i.e.} at $x=0^{-}$
to the values at the boundary to the vacuum at $x=-d$ using the Taylor
expansion: 
\begin{equation}
\hat{f}^{(F)}({\bf v}_{0},{{\bf r}_{0}}-{\bf r}_{{\bf d}})\approx \hat{f}%
^{(F)}({\bf v}_{0},{{\bf r}_{0}})-d\partial _{x}\hat{f}^{(F)}({\bf v}_{0},%
{\bf r}_{0})\,,  \label{Taylor}
\end{equation}
where ${{\bf r}_{0}}=(0,y,z)$ and ${\bf r}_{{\bf d}}=(-d,y,z)$ . Applying
general boundary conditions \cite{zaitsev} to the problem involved we
conclude that for a perfectly transparent interface the function $\hat{f}$
is continuous at the interface. At the boundary with the vacuum ($x=-d$) the
function $\hat{f}$ satisfies 
\begin{equation}
\hat{f}^{(F)}(v_{x},{\bf r}_{0}-{\bf r}_{{\bf d}})=\hat{f}^{(F)}(-v_{x},{\bf %
r}_{0}-{\bf r}_{{\bf d}})  \label{boundaryvacuum}
\end{equation}
Using Eqs. (\ref{eilenberger12}, \ref{Taylor}, \ref{boundaryvacuum}) and the
continuity of $\hat{f}$ at ${\bf r}={\bf r}_{0}$ the problem is reduced to the  solving of  the Usadel
equation in the superconductor with the following effective boundary
condition at the interface 
\begin{equation}
\eta D(\partial _{x}+d\partial _{y}^{2})\hat{f}_{0}({\bf r}_{0})+isgn\left(
\omega \right) \left( -\hat{V}\hat{f}_{0}+\hat{f}_{0}\hat{V}^{\ast }\right)
_{\!\!{\bf r}_{0}}\!\!\!\!=\!\!0  \label{boundaryinterface}
\end{equation}
where $\eta =v_{0}^{F}/v_{0}^{S}$ and $\hat{f}_{0}$ is the zero harmonics of
the function $\hat{f}$ in the superconductor. When deriving Eq. (\ref
{boundaryinterface}) we used the fact that the Usadel equation 
is applicable in the $S$- layer at distances down to the mean free path $l$
and extrapolated its solution to the interface. Only first two spherical
harmonics $\hat{f}^{(s)}\approx \hat{f}_{0}+{\bf v}_{{\bf 0}}\hat{{\bf f}}%
_{1}$ were kept in the derivation.

The linearized Usadel equation for the superconductor can be written in the
standard form 
\begin{equation}
D\nabla ^{2}\hat{f}_{0}-2|\omega |\hat{f}_{0}-2\Delta (x)\sigma _{y}=0
\label{a10}
\end{equation}

 The general solution of Eq. (\ref{a10}) with the boundary condition, Eq. (\ref{boundaryinterface}), and $\hat{V}$ from Eq. (\ref
{eilenberger12}) can be written as
\begin{equation}
\hat{f}_{0}({\bf r},\omega )=\alpha _{\omega }(x)\sigma _{x}e^{-i\sigma
_{x}Qy}+\beta _{\omega }(x)i\sigma _{y\,,}  \label{ansatz2}
\end{equation}
where $\alpha _{\omega }(x)=C_{\omega }\exp \left( -\sqrt{Q^{2}+\frac{%
2|\omega |}{D}}x\right) $ and $\beta _{\omega }(x)=-i\frac{\Delta (x)}{%
|\omega |}+B_{\omega }\exp \left( -\sqrt{\frac{2|\omega |}{D}}x\right) $.
Eq. (\ref{ansatz2}) is applicable at distances much smaller than $\xi \left(
T\right) $, where the solution for $\Delta $
can be approximated by a linear function. One can check using the
self-consistency Eq. (\ref{selbstconsist}) that the relative correction to $%
\Delta $ coming from the exponentially decaying part of  Eq. (%
\ref{ansatz2}), is of the order $\left( \ln \frac{\omega _{D}}{T_{c}}\right)
^{-1}$, where $\omega _{D}$ is the Debye frequency, and we neglected it. The
coefficients $C_{\omega }$ and $B_{\omega }$ can be now determined from Eq. (%
\ref{boundaryinterface}). Using the condition $\check{g}^{2}=1$ and Eq. (\ref
{energiaMS}) we can find the energy $\Omega _{M/S}$. Introducing the
dimensionless parameters: 
\begin{equation}
a^{2}\equiv \frac{2h^{2}d^{2}}{DT_{c}\eta ^{2}}\text{, \ }q^{2}\equiv \frac{%
DQ^{2}}{2T_{c}}\text{, \ \ }\widetilde{\Omega }\equiv \frac{\Omega }{\nu
_{F}\Delta _{0}^{2}}\sqrt{\frac{2T_{c}}{D}}  \label{a100}
\end{equation}
and using Eq. (\ref{orderparameter}) one obtains 
\begin{eqnarray}
\widetilde{\Omega }_{M/S} &=&\frac{\pi }{2}F_{3/2,1}K^{2}|\tau |+\sqrt{2}%
F_{2,1}K\left( 1-K^{2}\right) |\tau |^{3/2}  \nonumber \\
&&+\pi ^{-1}F_{5/2,1}\left( 1-K^{2}\right) ^{2}|\tau |^{2},  \label{a11} \\
F_{m,l} &=&\eta \frac{4a^{2}}{\pi ^{3/2-m}}\sum_{n>0}\alpha _{n}^{-m}\left[ 
\sqrt{\alpha _{n}\left( \alpha _{n}+q^{2}\right) }+a^{2}\right] ^{-l} 
\nonumber
\end{eqnarray}
where $\alpha _{n}=\pi (2n+1)$ and $\nu _{F}$ is the density of states in
the ferromagnet.The total energy is given by $\widetilde{\Omega }=\widetilde{\Omega }_{M}+%
\widetilde{\Omega }_{S}+\widetilde{\Omega }_{M/S}$,  Eqs. (\ref{magnetic}, \ref{energysuper}, \ref{a11}) and is  a functions of two parameters, $K$ and $q$, that should be
determined from the conditions $\partial \widetilde{\Omega }/\partial
K=\partial \widetilde{\Omega }/\partial q=0$. The parameter $q$ is in fact
the order parameter for the CF state. Close to the CF-F transition this
parameter is small and one can expand the energy $\widetilde{\Omega }_{M/S}$%
, Eq. (\ref{a11}), in $q^{2}$. As concerns the value $K_{0}$ at the minimum,
it can be found near the transition minimizing $\widetilde{\Omega }_{M/S}$
at $q=0$. As a result, the first terms of the expansion of the energy $%
\widetilde{\Omega }$ in $q^{2}$ near the CF-F transition can be written as

\begin{equation}
\scriptstyle 
\begin{array}{l}
\widetilde{\Omega }\approx \widetilde{\Omega }_{s}(K_{0})+\widetilde{\Omega }%
_{M/S}(K_{0},q=0)- \\ 
-\frac{q^{2}}{2}\left[ \frac{\pi }{2}F_{3/2,2}K_{0}^{2}|\tau |+\sqrt{2}%
F_{2,2}K_{0}(1-K_{0}^{2})|\tau |^{\frac{3}{2}}+\right. \\ 
\left. +\pi ^{-1}F_{5/2,2}\left( 1-K_{0}^{2}\right) ^{2}|\tau |^{2}-2\lambda 
\right] _{q=0}
\end{array}
\label{a13}
\end{equation}

One can check that the term proportional to $q^{4}$ is positive, which means
that the CF-F transition is of the second order. This is in contrast to the
conclusion of Ref. \cite{buzdin}. The parameter $\lambda $ in Eq. (\ref{a13}%
) is 
\begin{equation}
\lambda \equiv \frac{Jd}{\nu \sqrt{2T_{c}D^{3}}}\frac{7\zeta (3)}{2\pi ^{2}}
\label{a14}
\end{equation}
According to the Landau theory of phase transitions the transition from the
F state ($q=0$) to the CF state ($q\neq 0$) should occur when the
coefficient in the second-order term turns to zero. The phase diagram for
the variables $h$ and $J $, Eqs. (\ref{a100}, \ref{a14}), is
represented in Fig.1. The curves are plotted for different values of $|\tau |
$. The function $\widetilde{\Omega }(q)$ has only one minimum at $q_{0}$ continuously going  to zero as the system approaches the transition point.This demonstrates that the transition is of second order. Not close to the transition point $Q\sim \bar{\xi}^{-1}$.      

The stiffness $J$ for materials like $Fe$ and $Ni$ is $\approx 60K/$\AA .
Using the data for Nb $T_{c}\!=\!10$\AA , $v_{F}\!=\!1,37.10^{8}$cm/s, setting $%
l\!=\!100 $\AA , $d\!=\!10$\AA , and $h\!=\!10^{4}$K, which is proper for iron, and
assuming that the Fermi velocities and energies of the ferromagnet and
superconductor are close to each other we obtain $a\approx 25$ and $\lambda
\sim 6.10^{-3}$.  It is clear from Fig.1 that the CF state  is hardly possible in the $Fe/Nb$ structure studied in  
\cite{muhge}.

How can one explain the decay of the average magnetic moment below $T_{c}$
observed in that work? This can be understood if one assumes that there were
``islands'' in the magnetic layers with smaller values of $J$ and/or $h$. A
reduction of these parameters in the multilayers $Fe/Nb$ is not unrealistic
because proximity to $Nb$ leads to formation of non-magnetic ``dead'' layers 
\cite{zabel}, and can affect the parameters of the ferromagnetic layers,
too. If the CF state were realized only on the islands, the average magnetic
moment would be reduced but remain finite, which would correlate with the
experiment \cite{muhge}. One can also imagine  islands   very
weakly connected to the rest of the layer, which would lead to smaller
energies of a non-homogeneous state.

Another possibility to observe the CF state would be to use
multilayers with a weaker ferromagnet. A good candidate for this purpose
might be $Gd/Nb$. The exchange energy $h$ in $Gd$ is $h\approx 10^{3}K$ and
the Curie temperature and, hence, the stiffness $J$ is $3$ times smaller
than in $Fe$. So, one can expect $a\approx 2.5$ and $\lambda \approx
2.10^{-3}$. Using Fig.1 we see that the CF phase is possible for these
parameters. One can also considerably reduce the exchange energy $h$ in $%
V_{1-x}Fe_{x}/V$ multilayers \cite{aarts} varying the alloy composition.
Hopefully, the measurements that would allow to check the existence of the CF
phase in these multilayers will be performed in the nearest future.


\begin{figure}
\epsfysize = 5.5cm
\vspace{0.2cm}
\centerline{\epsfbox{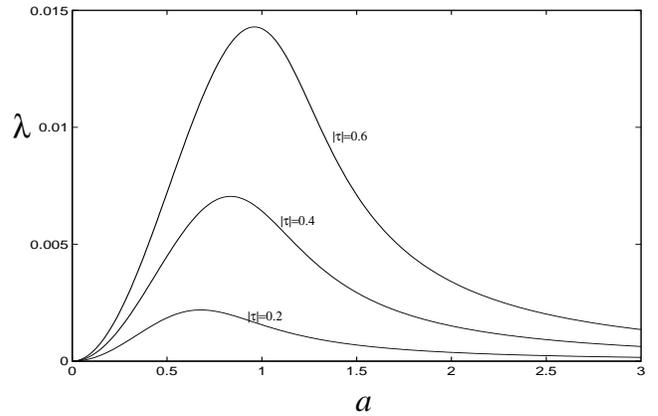
 }}
\vspace{0.2cm}
\caption{Phase diagrams ($\protect\lambda ,a$). The area above (below) the
curves corresponds to the F (CF) state }
\end{figure}

In conclusion, we studied a possibility of the CF state in ($S/F$)
 multilayers. We  derived a phase diagram that
allows to make definite predictions for real materials.

We are grateful to I.A. Garifullin for numerous discussion of experiments and to D. Taras-Semchuck and F.W.J. Hekking  for helpful discussions.
F.S.B. and K.B.E. thank SFB 491 {\it Magnetische Heterostrukturen }for a
support. The work of A.I.L. was supported by the NSF grant DMR-9812340 and
the A.v. Humboldt Foundation.


\end{document}